\documentclass[runningheads]{llncs}
\usepackage{graphicx}
\usepackage{amsmath, calc}
\usepackage[margin=1.5in]{geometry}

\usepackage[misc,geometry]{ifsym}

\usepackage{xcolor}

\begin{document}

\title{Our Nudges, Our Selves: Tailoring Mobile User Engagement Using Personality}
\titlerunning{Our Nudges, Our Selves: Tailoring Mobile User Engagement Using Personality}
%

\author{Nima Jamalian\inst{1}\and
Marios Constantinides\textsuperscript{(\Letter)}\inst{2} \and
Sagar Joglekar\inst{2} \and 
Xueni Pan\inst{1} \and 
Daniele Quercia\inst{2}}

\authorrunning{Jamalian et al.}
%

\institute{Goldsmiths, University of London, London, UK \email{\{n.jamalian, x.pan\}@gold.ac.uk}\\ \and
Nokia Bell Labs, Cambridge, UK
\email{\{marios.constantinides, sagar.joglekar, daniele.quercia\}@nokia-bell-labs.com}\\}

\maketitle              
\begin{abstract}
To increase mobile user engagement, current apps employ a variety of behavioral nudges, but these engagement techniques are applied in a one-size-fits-all approach. Yet the very same techniques may be perceived differently by different individuals. To test this, we developed HarrySpotter, a location-based AR app that embedded six engagement techniques. We deployed it in a 2-week study involving 29 users who also took the Big-Five personality test. Preferences for specific engagement techniques are not only descriptive but also predictive of personality traits. The Adj. $R^2$  ranges from 0.16 for conscientious users (encouraged by competition) to 0.32 for neurotic users (self-centered and focused on their own achievements), and even up to 0.61 for extroverts (motivated by both exploration of objects and places). These findings suggest that these techniques need to be personalized in the future.

\end{abstract}

\section{Introduction} 
\label{sec:introduction}
User engagement is crucial for the success of mobile apps, especially in modern Internet companies~\cite{lalmas2014measuring}. Mobile apps employ various techniques to capture users' attention and increase their engagement. For example, Foursquare introduced game mechanics to enhance engagement~\cite{lindqvist2011m}. Users could check-in at venues and inform their friends about their location. However, since not all friends may use the app, incentivizing early adopters became vital for the app's success. To motivate early adopters, Foursquare introduced badges, appealing to their desire for status. By default, the app shared this activity on social media platforms like Twitter, creating a sense of accomplishment and effectively engaging users. In general, engagement strategies, such as badges and rewards~\cite{brauer2019badges,van2019collecting}), encompass various mechanisms to increase user engagement~\cite{deterding2012gamification}. However, most current mobile apps follow a one-size-fits-all approach~\cite{adexchanger,rula2014no}, where all users are exposed to the same engagement techniques.

While users' personality has been extensively studied in various domains and linked to diverse aspects including online browsing behavior~\cite{bachrach2012personality,kosinski2014manifestations,quercia2011our} and patterns of behavior collected with smartphones~\cite{stachl2020predicting}, limited research has investigated its influence on mobile user engagement. To explore the relationship between personality and engagement strategies, we developed and deployed a location-based Augmented Reality (AR) mobile app called HarrySpotter, which incorporates six engagement techniques.

In this study, we made three sets of contributions. First, we developed HarrySpotter, a location-based app that enables users to annotate real-world objects. We conducted a two-week in-the-wild study involving 29 participants, resulting in a collection of 503 annotated objects. Second, we analyzed the engagement techniques chosen by users when capturing these objects and examined their correlation with their Big-Five personality traits. Our findings revealed that competition-based techniques discouraged agreeable users but encouraged conscientious users. Techniques promoting exploration of objects and places were particularly appealing to extroverts and individuals open to new experiences. Additionally, techniques focusing on personal achievements were found to motivate neurotic users. Lastly, we found that these preferences for specific engagement techniques not only correlated with personality traits but also had predictive value. The Adj. $R^2$ values ranged from 0.16 for conscientious users to 0.32 for neurotic users and as high as 0.61 for extroverts.
\section{Related Work}
\label{sec:related}
User engagement is a critical factor in the success of various digital experiences, including websites, mobile apps, and online platforms~\cite{lalmas2014measuring}. It refers to the level of involvement, interaction, and interest that users have with a product or service~\cite{o2008user}, resulting in increased engagement. Similarly, mobile user engagement is described by the level of engagement users 
have with mobile apps on their smartphones or tablets. Factors such as intuitive user interface design, personalized content delivery, and interactive features play a significant role in fostering mobile user engagement~\cite{sutcliffe2016designing,leiras2017mobile}. Push notifications, in-app messaging, and social sharing features also contribute to enhancing mobile user engagement~\cite{kim2018examining}.

Gamification techniques also play a crucial role in fostering engagement~\cite{deterding2012gamification,hakulinen2015effect}. Gamification involves applying game elements and mechanics to non-game contexts to enhance engagement, motivation, and participation~\cite{deterding2012gamification}. It taps into people's natural inclination for competition, achievement, and recognition, making it a powerful tool for motivating and incentivizing users~\cite{hamari2014does}. By incorporating game-like features such as points, badges, leaderboards, challenges, and rewards, mobile app developers can transform mundane tasks or activities into more enjoyable and immersive experiences. Rewards can take various forms, such as adding points or levels, to entice users to engage with an app to earn these rewards~\cite{nicholson2015recipe,lindqvist2011m}. Badges and leaderboards are also popular gamification elements, which were shown to boost motivation~\cite{brauer2019badges}.
Additionally, gamification strategies have been used to increase users' physical activity. For example, Althoff et al.~\cite{althoff2016influence} conducted a study on the impact of Pok\'emon GO~\cite{pokemongo}, an augmented reality (AR) location-based game, and found that the game led to a more than 25\% increase in users' physical activity.

While previous studies have explored the use of gamification strategies to engage mobile app users in a variety of tasks or games, the relationship between these strategies and a user's personality remains relatively unexplored. The purpose of this study is to investigate whether different individuals perceive the same gamification strategies differently.
\section{The Big-Five Personality}
\label{sec:hypotheses}
The Big-Five personality model assigns individuals scores~\cite{gosling2003very}, representing the main personality traits of Openness, Conscientiousness, Extroversion, Agreeableness, and Neuroticism. We hypothesized the relationship between these traits and our six engagement strategies, and summarized the positive and negative relationships of these hypotheses in Table~\ref{tab:questions}. \\

\noindent\textbf{\emph{Openness}} is associated with descriptive terms such as imaginative, spontaneous, and adventurous. Individuals high in Openness are more likely to try new methods of communication, including social networking sites or mobile apps. For example, studies have reported that individuals high in Openness tend to utilize a greater number of features that facilitate exploration in such technologies~\cite{ross2009personality}.

\noindent\textbf{\emph{Conscientiousness}} is associated with traits like ambition, resourcefulness, and persistence. Individuals high in Conscientiousness are less likely to engage in mobile content generation. They often view computer-mediated communication as a distraction from their daily tasks~\cite{amichai2010social}. However, when they do engage in such communication, they tend to approach it in a highly methodical and competitive manner. Their motivation is often driven by a desire for positive competition~\cite{halko2010personality}.

\noindent\textbf{\emph{Extraversion}} is associated with descriptive terms such as sociability, activity, and excitement seeking.  Individuals high in Extraversion typically prefer face-to-face interactions and are less inclined to utilize social networking sites or mobile apps. However, if they do join such platforms, they often participate in multiple groups, contribute content, and are motivated by the positive aspect of exploration as a means of social stimulation~\cite{phillips2006personality}.

\noindent\textbf{\emph{Agreeableness}} is associated with descriptive terms such as trusting, altruistic and tender-minded. Individuals high in Agreeableness, who are less competitive~\cite{halko2010personality} and less likely to share content~\cite{amichai2010social}, are more likely to be negatively motivated by rewards or competition.

\noindent\textbf{\emph{Neuroticism}} is associated with descriptive terms such as emotional liability and impulsiveness. Individuals with high levels of Neuroticism exhibit diverse behaviors across different media platforms. They tend to use the Internet and mobile apps as a means to alleviate loneliness, share accurate personal information in anonymous online forums (e.g., chat rooms), exercise control over their shared information on mobile devices~\cite{butt2008personality}, and focus on their own achievements in positive ways~\cite{lane2012influence}.

\begin{table*}[t]
\centering
\caption{Question statements assessing HarrySpotter's six engagement strategies. Positive and negative signs indicate the association of these engagement strategies with each personality trait as found in prior literature, and empty cells indicate that no reference has been found. O: Openness; C: Conscientiousness; E: Extraversion; A: Agreeableness; and N: Neuroticism. }
\resizebox{\textwidth}{!}
{\begin{tabular}{l l l l l l l} 
 \hline
 \textbf{Strategy} & \textbf{Question Statement} & \textbf{O} & \textbf{C} & \textbf{E} & \textbf{A} & \textbf{N} \\  \hline
 Q1(Point Rewards) & I pay attention to others' spell energy scores. &  &  &  & - & \\
 Q2(Place Rewards) & I am proud of my mayorships. &  &  &  & - & +\\ 
 Q3(Game with Yourself) & When I play the game, I feel I am representing my house. &  &  &  &  & + \\
 Q4(Social Connection) & With HarrySpotter, I track the competition among the four houses. &  & + &  & -  &\\
 Q5(Object Discovery) & HarrySpotter motivated me to discover new objects. & + &  & + &  &\\
 Q6(Place Discovery) & HarrySpotter motivated me to visit new places.  & + &  & + &  &\\
    \hline
\end{tabular}
}
\label{tab:questions}
\end{table*}
\section{HarrySpotter}
\label{sec:mobilegame}

\begin{figure*}[t]
\centering
\includegraphics[width=0.72\linewidth]{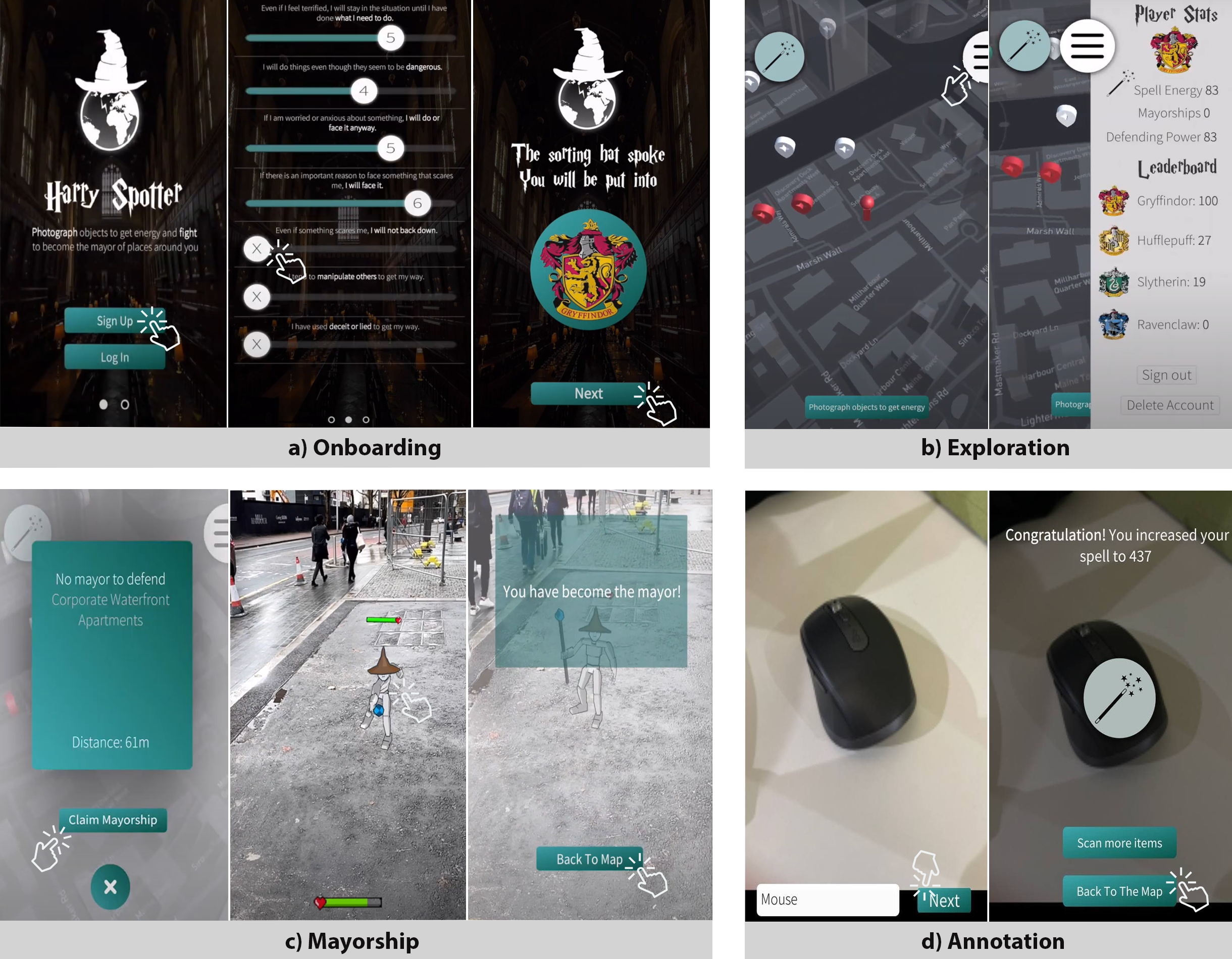}
\caption{HarrySpotter's gameplay elements: \emph{a) Onboarding}: set up account and complete the sorting hat quiz; \emph{b) Exploration}: location-based features (e.g., keeping track of mayorship around you), and leaderboard; \emph{c) Mayorship}: AR experience for claiming mayorship (a mage would appear to challenge the user for mayorship); and \emph{d) Annotation:} capture and annotate objects.}
\label{fig:strategies}
\end{figure*}

We developed a mobile app called HarrySpotter, which incorporates gameplay elements inspired by the popular Harry Potter series. Authored by J.K. Rowling, the Harry Potter series revolves around the adventures of a young wizard named Harry Potter, his friends, and their quest to defeat the dark wizard Lord Voldemort. Our app draws inspiration from the series' concept of affiliation through four houses, namely Gryffindor (known for courage and bravery), Hufflepuff (emphasizing hard work and patience), Ravenclaw (highlighting intelligence and learning), and Slytherin (representing ambitions and cunning). HarrySpotter was developed using the Unity game engine for both Android and iOS platforms. The app uses the Mapbox SDK for location-based features, and Vuforia SDK to deliver an augmented reality experience, particularly during the process of claiming a mayorship.

HarrySpotter employs six strategies to engage users in the task of annotating objects: Point Rewards, Places Rewards, Game with Yourself, Social Connection, Object Discovery, and Place Discovery. These strategies were initially derived from the work of Lindqvist et al.~\cite{lindqvist2011m}, but were modified to align with our gameplay's requirements. For example, the effectiveness of badges in Lindqvist's study was evaluated using the question statement: ``I pay attention to the badges that others earn.'' In our case, to assess the effectiveness of point rewards (which function as a type of badge), we adapted the statement to: ``I pay attention to others' spell energy scores'' (Table\ref{tab:questions}). Additionally, as our gameplay includes two types of rewards---points and place rewards---we categorized them separately, resulting in a total of six strategies, contrasting with the five strategies described in~\cite{lindqvist2011m}. \\ 

\noindent 
\textbf{Point and Place Rewards}: Previous research has demonstrated the effectiveness of reward systems, such as points, in engaging users with mobile apps~\cite{brauer2019badges}. In HarrySpotter, users are rewarded with spell energy for annotating new objects (Figure~\ref{fig:strategies}d), which reflects their ability to claim mayorships of places. When a user annotates an object, the app compares the user-generated label (object name) with the label automatically detected by an image classifier running on our server. The classifier used is a deep-learning ResNet-162 model with a top 5\% accuracy of 94.2\% on ImageNet classes. The semantic distance between the user-generated label and the automatically detected label is computed using WordNet~\cite{miller1998wordnet}. If they match, the user receives extra spell energy. Additionally, the app tracks previously scanned object types and rewards the user when they scan a new type for the first time. However, if the user scans the same object type repeatedly, the reward amount decreases until it reaches the minimum of 10 points. This design choice ensures that the user's score does not reach zero and maintains a balance between engagement and avoiding penalization, such as providing incorrect labels or repetitive images.

\noindent \textbf{Game With Yourself}: Users have the option to play the game alone, engaging in various single-player elements such as object annotation, places, and challenging mayorships (Figure~\ref{fig:strategies}b-d). When it comes to mayorships, a user can become the mayor of a place. Subsequently, other users have the opportunity to visit that place and challenge the current mayor. This feature adds a competitive aspect to the game, even when playing individually.

\noindent\textbf{Social Connection}: Previous research has demonstrated that leaderboards are effective in enhancing user performance in various tasks~\cite{brauer2019badges}. In HarrySpotter, during the onboarding process, users respond to a series of questions inspired by the Harry Potter sorting hat quiz (Figure~\ref{fig:strategies}a) and are sorted into one of the Harry Potter houses~\cite{jakob2019science}. Through the leaderboard, we encourage users to actively participate in the game and contribute to their respective houses' efforts in claiming mayorships of different places.

\noindent \textbf{Object Discovery}: Enabling users to explore and discover new places or objects is a crucial aspect of location-based apps. Previous research has demonstrated that incorporating points of interest (similar to Pok\'emon GO) encourages users to engage with the app while on the move and at various locations~\cite{althoff2016influence}. In HarrySpotter, we motivate users to explore different locations by allowing them to become mayors of real places (Figure~\ref{fig:strategies}c). When a user is within the mayorship location range, an AR mage appears and challenges the user for mayorship. The user's spell energy (points) plays a significant role in their chances of claiming the mayorship. To strike the right balance, we set the mayorship range to an 80-meter radius based on empirical evidence. Lower ranges limited accessibility, while higher ranges diminished proximity and overall engagement. Through this strategy, we encourage users to discover and scan objects they may have overlooked in new locations, fostering exploration and engagement.

\noindent \textbf{Place Discovery.} Previous research on the motivations behind using location-based apps has revealed that users are driven by their curiosity to obtain information about specific points of interest~\cite{lindqvist2011m}. This curiosity acts as an incentive for users to actively pursue becoming the mayor of those places. In HarrySpotter, when a user successfully claims mayorships of places, their map visually represents a sense of territorial ownership (Figure~\ref{fig:strategies}b). For example, if a user belongs to Hufflepuff house and becomes a mayor, their map pins will be displayed in yellow, symbolizing their affiliation with Hufflepuff.
\section{User Study}
\label{sec:userstudy}

\noindent
\textbf{Participants and Ethical Considerations.} We deployed HarrySpotter in a 2-week study with 29 users (13 female), aged between 18-49 years (median = 34). To be eligible for the study, participants were required to own an Android or iOS smartphone and be located in London, UK. In compliance with GDPR and the Data Protection Act, all individual user data were anonymized to ensure the privacy and confidentiality of the participants. The study was approved by the Ethics Committee of Goldsmiths, University of London.

\noindent \textbf{Procedure.} All participants underwent a pre-screening process where we collected demographic information and obtained the unique identifier of their device for generating the app download link. After installing the app, participants were prompted to grant access to the camera and location. Basic instructions were provided on how to use the app, such as annotating objects, with no specific guidelines on what to annotate or how frequently. To maintain study integrity, no information regarding the relationship between personality and engagement techniques was revealed to the participants.

\noindent \textbf{Materials and Apparatus.} At the end of the study, we administered a 6-item questionnaire (Table~\ref{tab:questions}) and the 10-item TIPI personality questionnaire~\cite{gosling2003very}. The 6-item questionnaire included statements derived from~\cite{lindqvist2011m} and had previously been validated in the context of the Foursquare app to assess users' motivations for engagement. Participants rated both questionnaires using a 7-point Likert scale (1: Strongly Disagree; 7: Strongly Agree).

\noindent
\textbf{Self-reports and Big-Five personality traits.} We coded the Likert-scale answers to the 6-item questionnaire and the TIPI~\cite{gosling2003very}. On average, our participants scored as follows on a 1-7 scale: average in Openness ($\mu\textrm{=5.14}$, $\sigma\textrm{=0.8}$), high in Conscientiousness ($\mu\textrm{=5.12}$, $\sigma\textrm{=0.85}$), average in Neuroticism ($\mu\textrm{=4.59}$, $\sigma\textrm{=0.85}$), average in Agreeableness ($\mu\textrm{=4.5}$, $\sigma\textrm{=0.71}$), and low in Extraversion ($\mu\textrm{=4.43}$, $\sigma\textrm{=0.94}$). These trait distributions aligned with the normative personality values derived from a large sample of the U.S. population~\cite{soto2011age}.

\noindent\textbf{Annotations.} Each annotation in our study involved storing the raw image and its corresponding label in a database. To ensure data quality, we implemented checks for image duplication and semantic correctness. To prevent duplication, we utilized FAISS~\cite{johnson2019billion}, a framework for indexing images based on visual similarity. This allowed us to retrieve the most visually similar images for comparison. We penalized scores for each annotation based on visual similarity to the user's previously uploaded images. For instance, if an image closely resembled a previously captured one, the user would not receive a reward in the form of spell energy. To assess semantic correctness, we first subjected the uploaded image to an off-the-shelf object detector~\cite{cheng2019panoptic}. We then calculated the WordNet semantic distance~\cite{miller1998wordnet} between the detected label and the user-generated label. The awarded spell energy was proportional to the semantic similarity, discouraging grossly inaccurate or garbled labels. Regarding annotation quantity, we recorded the total number of annotations $n_k$ uploaded by each user $k$ along with their respective images. For annotation quality, three independent annotators rated each annotation on a 1-5 Likert scale, with 5 indicating a perfect match between the image and the user-generated label. For example, if an image depicted a ``computer mouse'' and the user's label was ``mouse,'' the annotator would assign a score of 5. To ensure reliable results, we calculated a Fleiss kappa score of 0.57, indicating moderate to good agreement among the three annotators. We compiled a set of $n$ images $I$ annotated by each user $k$ as ${I_1, I_2, ..., I_n}$. The quality score for user $k$ was determined by the median of the quality scores assigned to their $n$ annotated images.

Before using the six self-reports and the quantity and quality of annotation metrics into our regression models, we conducted a Shapiro-Wilk test for normality. As the eight variables exhibited skewed distributions, we applied a log transformation to them. Among the five personality traits, only Extraversion showed a slight skewness, so we also applied a log transformation to it.

\section{Results}
\label{sec:results}
To ease the interpretation of our results, we applied a min-max transformation to scale our variables within the range of [0-100]. We first examined the pairwise correlation among personality traits, the six self-reports, and the quantity and quality of annotations metrics. We found that neurotic users discovered fewer objects ($\textrm{r=-0.37, p\textless0.1}$). Additionally, these users tended to take pride in their mayorships ($\textrm{r=0.44, p\textless0.05}$) and their affiliation with their respective Harry Potter house ($\textrm{r=0.47, p\textless0.05}$). Conversely, Extroverts and those high in Openness liked to discover new objects ($\textrm{r=0.44, p\textless0.05}$ and $\textrm{r=0.39, p\textless0.05}$, respectively) and new places ($\textrm{r=0.38, p\textless0.05}$ and $\textrm{r=0.35, p\textless0.1}$, respectively). 

\begin{table*}[ht!]
\centering
\caption{Linear regressions that predict the Big-Five personality traits from the six self-reports and the quantity and quality of annotations. Significant predictors with $p$ values $<$ .05 are marked in bold. The most predictable personality trait was Extraversion ($M_{E}$), while the least predictable was Conscientiousness ($M_{C}$).}
      \begin{tabular}{lccc}
        \hline
        $\boldsymbol{M_{O}}$: $\boldsymbol{\emph{Adj $R^2$} = 0.28}$, Durbin-Watson = 1.96, AIC = 0.47 & & &  \\
        \hline
        Predictor & $\beta$ & std. error & $p$-value \\
        \hline
        Intercept & 0.42 & 0.13 & 0.005\\
        Q1(Point Rewards) & -1.08 & 0.36 & \textbf{0.01}\\ 
        Q2(Place Rewards) & 0.42 & 0.35 & 0.23\\ 
        Q5(Object Discover) & 0.84 & 0.27 & \textbf{0.01}\\ 
        \hline
        $\boldsymbol{M_{C}}$: $\boldsymbol{\emph{Adj $R^2$} = 0.16}$, Durbin-Watson = 1.15, AIC =  -7.48 & & &  \\
        \hline
        Predictor & $\beta$ & std. error & $p$-value \\
        \hline
        Intercept & 0.57 & 0.12 & 0.00\\
        Q3(Game with Yourself) & 0.59 & 0.22 & \textbf{0.01}\\ 
        Q6(Place Discovery) & -0.48 & 0.23 & \textbf{0.04}\\
        \hline
        $\boldsymbol{M_{E}}$: $\boldsymbol{\emph{Adj $R^2$} = 0.61}$, Durbin-Watson = 1.88, AIC = -14 & & &  \\
        \hline
        Predictor & $\beta$ & std. error & $p$-value \\
        \hline
        Intercept & 0.48 & 0.19 & 0.02\\
        Q2(Place Rewards) & -0.60 & 0.22 & \textbf{0.01}\\ 
        Q4(Social Connection) & -0.72 & 0.22 & \textbf{0.00}\\ 
        Q5(Object Discovery) & 0.53 & 0.22 & \textbf{0.02}\\ 
        Q6(Place Discovery) & 0.99 & 0.28 & \textbf{0.00}\\
        Quantity & -0.37 & 0.12 & \textbf{0.00}\\
        Quality & 0.18 & 0.16 & 0.27\\
        \hline
        $\boldsymbol{M_{A}}$: $\boldsymbol{\emph{Adj $R^2$} = 0.21}$, Durbin-Watson = 1.84, AIC = -12.86 & & &  \\
        \hline
        Predictor & $\beta$ & std. error & $p$-value \\
        \hline
        Intercept & 0.57 & 0.11 & 0.00\\
        Q1(Point Rewards) & -0.45 & 0.24 & 0.07\\  
        Q3(Game with Yourself) & 0.21 & 0.21 & 0.31\\
        Q5(Object Discovery) & -0.37 & 0.24 & 0.13\\
        Q6(Place Discovery) & 0.71 & 0.29 & \textbf{0.02}\\  
        \hline
        $\boldsymbol{M_{N}}$: $\boldsymbol{\emph{Adj $R^2$} = 0.32}$, Durbin-Watson = 2.79, AIC = -3.68 & & &  \\
        \hline
        Predictor & $\beta$ & std. error & $p$-value \\
        \hline
        Intercept & 0.36 & 0.14 & 0.02\\
        Q2(Place Rewards) & 0.37 & 0.26 & 0.17\\  
        Q4(Social Connection) & 0.58 & 0.27 & \textbf{0.04}\\
        Q6(Place Discovery) & -0.58 & 0.29 & 0.06\\
        Quantity & -0.32 & 0.14 & \textbf{0.03} \\
        \hline
      \end{tabular}
      \label{tab:regression}
\end{table*}

Considering these significant correlations, one might wonder whether it is possible to predict users' personality traits based on their self-reports and the quantity and quality annotation metrics. Using these metrics as predictors, we fitted five linear regression models (Table~\ref{tab:regression}) to predict the Big-Five personality dimensions and determined the best set of predictors using the stepAIC function~\cite{zhang2016variable}. Overall, our findings indicate that predicting certain personality dimensions, such as Extraversion (Adj. $R^2$ = 0.61), was relatively easier compared to others like Conscientiousness (Adj. $R^2$ = 0.16).

As expected, users with non-competitive traits (high in Agreeableness) demonstrated a lack of motivation for competing with others ($\beta_\textrm{(Q1(Point Rewards))}\textrm{=-0.45}$). On the other hand, conscientious users, known for their organizational skills, exhibited motivation for competition ($\beta_\textrm{(Q1(Point Rewards))}\textrm{=0.59}$) among the four houses but moderately predicted their personality trait (Adj. $R^2$\textrm{=0.16}). Individuals open to new experiences and extroverts were primarily motivated by the discovery of new objects ($\beta_\textrm{(Q5(Object Discovery))}\textrm{=0.84}$ and $\beta_\textrm{(Q5(Object Discovery))}\textrm{=0.53}$, respectively). Moreover, individuals open to new experiences did not find motivation in competing with others ($\beta_\textrm{(Q1)}\textrm{=-1.08}$), while extroverts were not motivated by mayorships ($\beta_\textrm{(Q2(Place Rewards))}\textrm{=-0.60}$) or representing their own house ($\beta_\textrm{(Q1(Point Rewards))}\textrm{=-0.72}$). Emotionally unstable users (neurotics) found motivation in representing their own house ($\beta_\textrm{(Q4(Social Connection))}\textrm{=0.58}$) but not in the discovery of new places ($\beta_\textrm{(Q6(Place Discovery))}\textrm{=-0.58}$), leading them to discover fewer objects ($\beta_\textrm{(Quantity)}\textrm{=-0.32}$).
\section{Discussion and Conclusion}
\label{sec:discussion}
Mobile user engagement is commonly pursued through a range of techniques; however, there is a tendency to apply these techniques uniformly to all users, adopting a one-size-fits-all approach. To investigate the possibility that individuals may perceive engagement techniques differently, we created HarrySpotter, a location-based augmented reality (AR) app that enables users to annotate real-world objects using six distinct engagement techniques. By deploying HarrySpotter and analyzing data from 29 users, we found that agreeable users were not motivated by competition, while conscientious users were motivated by it to a greater extent. As expected, individuals open to new experiences and extroverts were motivated by exploration, while neurotics exhibited a stronger drive towards personal achievements. These preferences for specific engagement techniques also predicted personality traits to different extents (e.g., Extraversion with an Adj. $R^2$ of 0.61, while Conscientiousness with an Adj. $R^2$ of 0.16), suggesting that engagement strategies should be tailored to one's personality.

From a theoretical perspective, our work is situated within the domain of adaptive user interfaces (AUI). The effectiveness of an AUI hinges on the ability to construct and utilize individual user profiles, allowing for the delivery of personalized versions of the user interface~\cite{jameson2007adaptive,constantinides2015exploring,constantinides2018framework,constantinides2016user}. Building upon this foundation, we envision that our methodology could be employed to enhance user models with specific characteristics, such as personality traits. This, in turn, could facilitate the personalization of user interfaces in various contexts, such as advancing levels in a gamified app or completing tasks. From a practical standpoint, our findings can inform the design of personalized engagement strategies. To illustrate, let us consider the scenario of mobile crowdsourcing systems, where a one-size-fits-all approach has proven ineffective in engaging users for specific tasks, such as object annotation~\cite{rula2014no}. By incorporating brief personality questionnaires, for example during account setup (e.g., the TIPI~\cite{gosling2003very} questionnaire, which can be completed in a minute), mobile developers can implement in-app mechanisms to dynamically infer personality traits, thereby adapting engagement strategies based on users' interactions.

Our work has three limitations that warrant further research efforts. Firstly, our findings are specific to the HarrySpotter game and this particular cohort. Future studies could extend our methodology to different types of mobile apps. For example, developing tailored strategies for Conscientiousness could enhance prediction accuracy by incorporating logging mechanisms for organized individuals. Secondly, the slight skew in Extraversion may be due to self-selection bias, as introverted individuals are less likely to engage with such apps. Future studies should replicate our methodology with larger and culturally diverse populations. Lastly, while our two-week study provided ample data, longer deployments can explore user retention and preferences more comprehensively, thereby enhancing our understanding of personalized engagement strategies.

%
%
%
\bibliographystyle{splncs04}
\bibliography{main}

\end{document}